\documentclass[]{aa} % for a referee version
\begin{document}
\outer\def\gtae {$\buildrel {\lower3pt\hbox{$>$}} \over 
{\lower2pt\hbox{$\sim$}} $}
\outer\def\ltae {$\buildrel {\lower3pt\hbox{$<$}} \over 
{\lower2pt\hbox{$\sim$}} $}
\newcommand{\ergscm} {ergs s$^{-1}$ cm$^{-2}$}
\newcommand{\ergss} {ergs s$^{-1}$}
\newcommand{\ergsd} {ergs s$^{-1}$ $d^{2}_{100}$}
\newcommand{\pcmsq} {cm$^{-2}$}
\newcommand{\ros} {\sl ROSAT}
\newcommand{\exo} {\sl EXOSAT}
\newcommand{\xmm} {\sl XMM-Newton}
\newcommand{\chan} {\sl Chandra}
\def\rchi{{${\chi}_{\nu}^{2}$}}
\def\uchi{{${\chi}^{2}$}}
\newcommand{\Msun} {$M_{\odot}$}
\newcommand{\Mwd} {$M_{wd}$}
\def\Mdot{\hbox{$\dot M$}}
\def\mdot{\hbox{$\dot m$}}
\input psfig.sty
%\newcommand{\deg} {$^{\circ}$}
%
%\thesaurus{}
\title{\sl{XMM-Newton} observations of AM CVn binaries}
\titlerunning{\sl{XMM-Newton} observations of AM CVn binaries}
\authorrunning{Ramsay et al} 

%\subtitle{}

\author{Gavin Ramsay \inst{1}, Pasi Hakala\inst{2}, Tom
Marsh$^{3}$, Gijs Nelemans$^{4,5}$, Danny Steeghs$^{6}$, \and Mark
Cropper$^{1}$}

\offprints{G. Ramsay}

\institute{
$^{1}$Mullard Space Science Laboratory, University College London,
Holmbury St Mary, Dorking, Surrey, RH5 6NT, UK.\\
$^{2}$Observatory, P.O. Box 14, FIN-00014 University of Helsinki,
Finland.\\
$^{3}$Department of Physics,
University of Warwick, Coventry, CV4 7AL, UK\\ 
$^{4}$Institue of Astronomy, University of Cambridge, Madingley Road, 
Cambridge, CB3 0HA, UK\\ 
$^{5}$Department of Astrophysics, Radboud University of Nijmegen, PO
Box 9010, 6500 Nijmegen, The Netherlands\\
$^{6}$Harvard-Smithsonian Center for Astrophysics, 60 Garden
Street, MS-67, Cambridge, MA 02138, USA\\ }

%\email{gtbr@mssl.ucl.ac.uk}

\date{}

\abstract{ We present the results of {\xmm} observations of four AM
  CVn systems -- AM CVn, CR Boo, HP Lib and GP Com.  Their light
  curves show very different characteristics. The X-ray light curves
  show no coherent pulsations, suggesting the accreting white dwarfs
  have relatively low magnetic field strengths. Their spectra were
  best modelled using a multi-temperature emission model and a strong
  UV component.  We find that CR Boo and HP Lib have X-ray spectra
  with abundances consistent with relatively low temperature CNO
  processed material, while AM CVn and GP Com show an enhancement of
  nitrogen. A large fraction of the accretion luminosity is emitted in
  the UV. We determine accretion luminosities of
  $\sim1.6\times10^{33}$ \ergss and 1.7$\times10^{31}$ \ergss for AM
  CVn and GP Com respectively. Comparing the implied mass transfer
  rates with that derived using model fits to optical and UV spectra,
  we find evidence that in the case of AM CVn, we do not detect a
  significant proportion of the accretion energy. This missing
  component could be lost in the form of a wind.  }

\maketitle

\section{Introduction}

Most Cataclysmic Variables (CVs) consist of a white dwarf accreting
material from a main sequence secondary star through Roche lobe
overflow. However, in some CVs (the AM CVn stars) the mass-donating
star is hydrogen deficient -- either a helium star or a white dwarf
(see Warner 1995 for a review). Their optical spectra display no
hydrogen features but instead are dominated by helium lines. Since
these short orbital period systems are very compact, they are expected
to be strong sources of gravitational radiation.

Currently there are $\sim$13 such systems known, with orbital periods
ranging from 5--65 mins. The longer period systems, such as GP Com and
CE 315, are expected to have lower mass transfer rates than the
shorter period systems. The nature of the two shortest period systems,
RX J0806+15 (a period of 5.4 min) and RX J1914+24 (9.5 min), remain
the subject of some controversy (eg Nelemans 2004), but do not show
evidence for an accretion disc which is typically seen in other AM CVn
systems.

AM CVn stars have been little studied in X-rays. In the {\sl Rosat}
all-sky survey only 3 were detected (Ulla 1995). Even for those
sources, there is a wide difference in the X-ray to optical ratios.
In the UV band at least some systems have been found to be bright (eg
van Teeseling, Beuermann \& Verbunt 1996). {\xmm} with its large
effective area X-ray telescopes and its optical/UV telescope, is
therefore an ideal satellite with which to obtain multi-wavelength
observations of these systems. This paper presents the results of
{\xmm} observations of 4 AM CVn systems: AM CVn, HP Lib, CR Boo and GP
Com. We relate their X-ray and UV properties to the longer period,
hydrogen accreting CVs.

\section{Observations}

{\xmm} has 3 broad-band X-ray detectors with medium energy resolution
and also a 30 cm optical/UV telescope (the Optical Monitor, OM: Mason
et al 2001). The X-ray instruments contain imaging detectors covering
the energy range 0.15--10keV. Currently the EPIC pn detector
(Str\"{u}der et al 2001) is better calibrated at lower energies
compared to the EPIC MOS detector (Turner et al 2001). Two high
resolution grating spectrometers (the RGS) are also on board. However,
apart from GP Com (whose RGS spectra are discussed in Strohmayer 2004)
the sources are too faint to obtain useful spectra. The observation
log is shown in Table \ref{log} where we show the mean X-ray and UV
count rates for each source.

The data were processed using the {\sl XMM-Newton} {\sl Science
  Analysis Software} (SAS) v6.0 and analysed in a similar manner to
that in Ramsay et al (2005). For GP Com, we used the EPIC MOS data
since the EPIC pn data were moderately piled up. With the exception of
GP Com, each target was observed in the OM using the fast mode and
UVW1 filter: this has an effective wavelength of 2910\AA\hspace{1mm}
and range of 2400--3400\AA. The fast mode data were analysed using the
SAS task {\tt omfchain}. GP Com was observed in the OM in full mode
and the $V$, $B$, $U$, UVW1 and UVW2 filters (effective wavelength
2120\AA).

\begin{table}
\begin{center}
\begin{tabular}{lcrrr}
\hline
Source & Date & Exp & EPIC  & UV \\
       &      & (ksec) & Ct/s & Ct/s \\
\hline
AM CVn & 2003-05-23 & 10.4 & \\
AM CVn & 2003-11-24 & 10.5 & 0.20 & 95 \\
CR Boo & 2003-12-28 & 20.0 & 0.12 & 35 \\
HP Lib & 2004-01-28 & 20.0 & 0.65 & 130 \\
GP Com & 2001-01-03 & 51.2 & 2.4 & 12.1 \\
\hline
\end{tabular}
\end{center}
\caption{The observation log for the sources discussed in this
paper. The exposure time is that of the EPIC pn detector. The count
rate refers to the mean count rate in the EPIC pn detector
(0.15--10keV) and the UVW1 filter (2400-3400\AA). The first
observation of AM CVn was affected by high solar particle background.}
\label{log}
\end{table}

\section{Light curves}

\begin{figure*}
\begin{center}
\setlength{\unitlength}{1cm}
\begin{picture}(13,10)
\put(-1.5,-0.8){\includegraphics{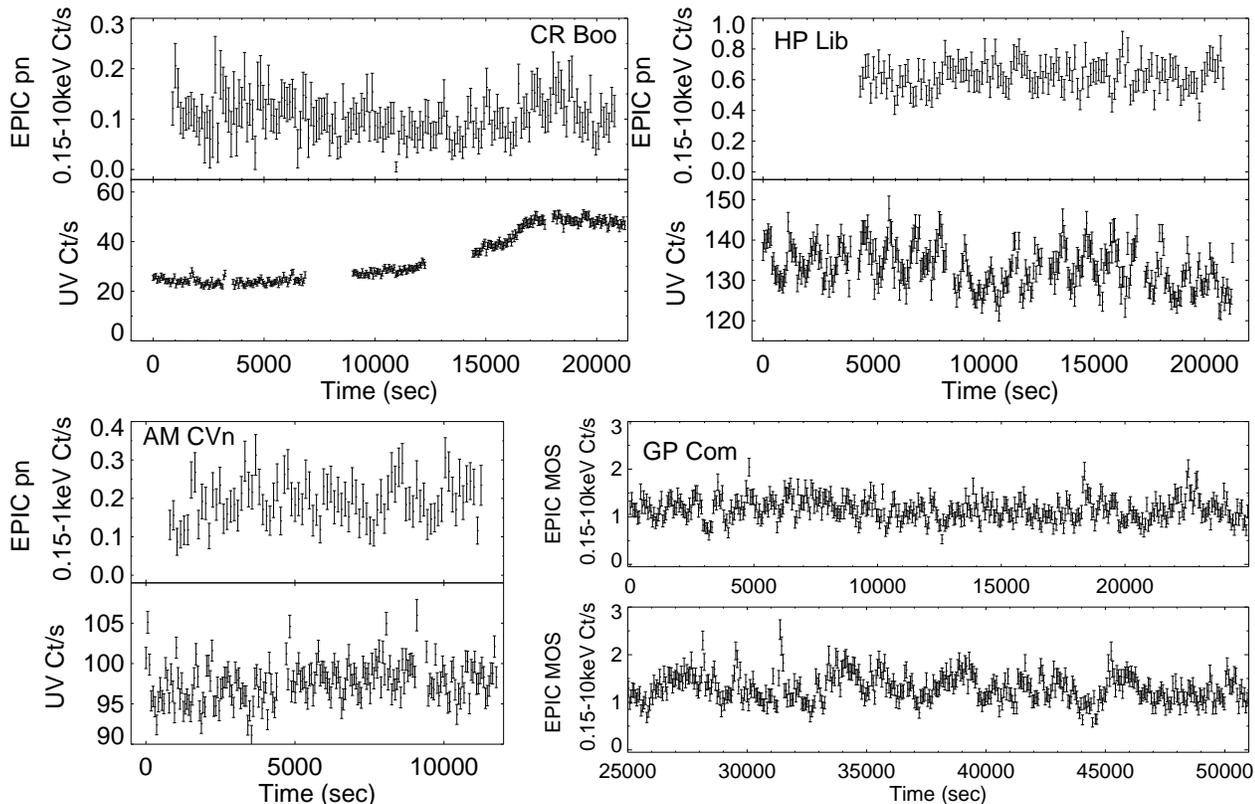}}
\end{picture}
\end{center}
\caption{The X-ray and UV light curves for our sample of AM CVn
systems. In the case of GP Com no fast mode UV data was obtained. The
OM data has time bins of 60 sec, the X-ray data of CR Boo, HP Lib and
AM CVn 120 sec and GP Com 60 sec.}
\label{lightcurves}
\end{figure*}

The X-ray and UV light curves for each source are shown in Figure
\ref{lightcurves}. In the UV, AM CVn shows little obvious coherent
modulation, while HP Lib shows the presence of a clear periodic
signal. In contrast, CR Boo shows a large variation over the course of
the observation, increasing from $\sim$25 ct/s to $\sim$50 ct/s. The
timescale is longer than the 1492 sec period reported by Provencal et
al (1997) but much shorter than the 46 day `long' super-cycle reported
by Kato et al (2000).  The X-ray light curves do not show strong
modulations such as that seen in the UV light curve of HP Lib.

A simple analysis of the X-ray light curves shown in Figure
\ref{lightcurves} using the Lomb Scargle algorithm or a DFT produces
many peaks in the corresponding power spectrum. However, red noise can
lead to the appearance of peaks in the power spectra which can be
mistaken for coherent modulation.  We use the method of Hakala et al
(2004) to determine the significance of peaks in power spectra.

We performed this analysis on the X-ray and UV data. We show the power
spectra of the X-ray light curves in Figure \ref{xpower}. In X-rays,
the spectra of HP Lib shows no peaks above the 99 percent
(=2.6$\sigma$) confidence interval. AM CVn has two peaks above
this confidence level, at $\sim$1631 sec and 338 sec, while CR Boo has 
two, at 742
sec and one at 15124 sec.  GP Com has 4 peaks which are just above the
99 percent confidence level: at 372, 682 sec, 2271 sec and 3951
sec. For all of these peaks, none are related to the known periods for
these systems. Further, the known periods in the systems do not
correspond to significant peaks in their power spectra. In summary,
although there are some peaks which are moderately significant (ie
above the 99 percent level), we do not consider them to be strongly
significant and therefore there is no strong evidence for coherent
modulation in the X-ray light curves.

\begin{figure}
\begin{center}
\setlength{\unitlength}{1cm}
\begin{picture}(6,17)
\put(-2.,18){\includegraphics{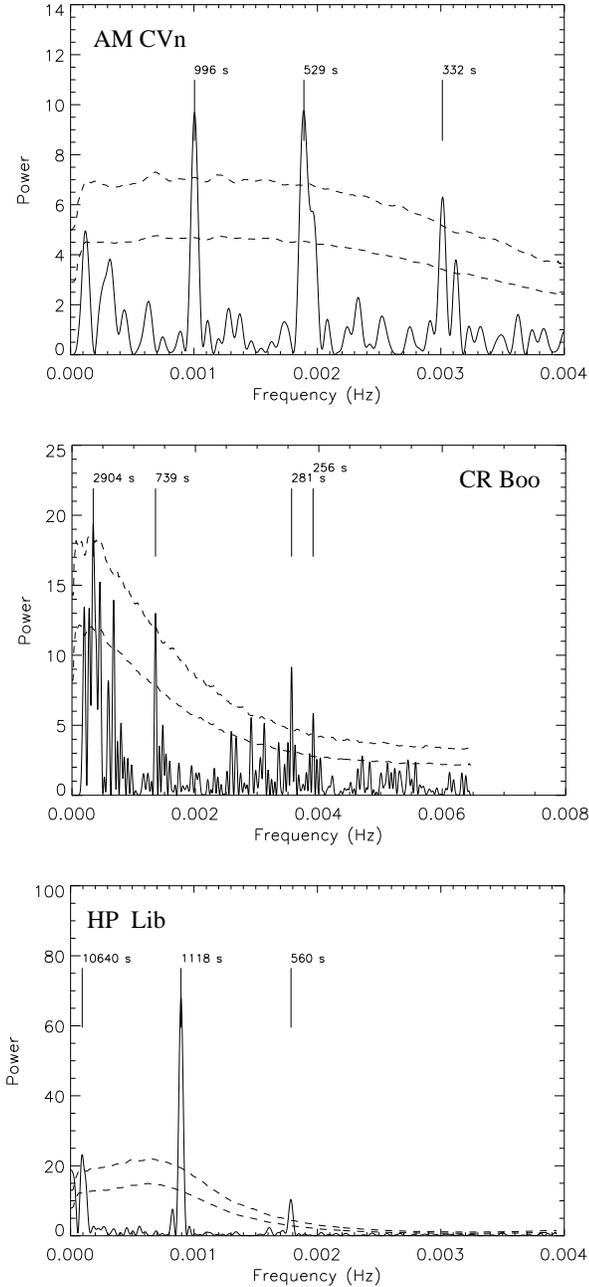}}
\end{picture}
\end{center}
\caption{The power spectra for the UV light curves of AM CVn, CR
Boo and HP Lib, together with the 95, 99 percent confidence
intervals based on our red noise analysis.}
\label{uvpower}
\end{figure}

The UV data of HP Lib shows an obvious modulation on a period of 1117
sec. This is similar to the period of the super-hump reported by
Patterson et al (2002) which varied between 1118.89--1119.14 sec in
their observations.  The power spectrum also shows a period near 560
sec which appears to be a harmonic of the main peak (Figure
\ref{uvpower}). The AM CVn data shows 3 periods above the 99 percent
confidence level: 996 sec, 529 sec and 332 sec. We estimate the error
on the longest period to be 11 sec: the 996 sec period is therefore 
not significantly different to the 
strongest period (1014$\pm$2 sec) detected in the UV by Solheim,
Provencal \& Sion (1997) and also the 1011.4 sec modulation seen in
the optical (Provencal 1995). In the case of CR Boo we fitted a
polynomial to remove the trend in the light curve. The peak with
greatest significance above the 99 percent confidence interval is a
peak at 281 sec.

We searched for a correlation between the intensity behaviour in the
X-ray and UV energy bands using the method of Hakala et al (2004).
Only HP Lib shows a marginal correlation between these light curves:
after we had removed the general trend in the UV light curve, there is
a broad peak in the correlation function between -450 to -200 sec,
implying the X-ray trails the UV light curve. However, the most
prominent X-ray peak, which occurs at 11500 sec, coincides in time
with the UV peak.

\begin{figure*}
\begin{center}
\setlength{\unitlength}{1cm}
\begin{picture}(12,11)
\put(-4,12.5){\includegraphics{xray_pds.ps}}
\end{picture}
\end{center}
\caption{The power spectra for the X-ray light curves of AM CVn, CR
Boo, GP Com and HP Lib, together with the 95 and 99  percent
significance intervals based on our red noise analysis.}
\label{xpower}
\end{figure*}

\section{Spectra}
\label{spec}

\begin{table}
\begin{center}
\begin{tabular}{lrrrr}
\hline
Source & cemekl & cemekl & cevmkl & cevmkl \\
       &  (Z=1)  & Z var  &  (CNO)  &  Z var     \\
\hline
HP Lib & 1.97 (213) & 1.60 (212) & 1.18
(212) & 1.16 (213) \\
CR Boo & 1.15 (72) & 1.15 (71) & 1.05 (72) &
1.05 (65) \\
GP Com &  14.9  (160) & 6.35 (159) & 7.77
(160) & 1.32 (153)\\
AM CVn & 2.54 (35) & 2.20 (34) & 1.31 (35) &
1.02 (34) \\
\hline
\end{tabular}
\end{center}
\caption{The fits (\rchi, dof) to the {\xmm} EPIC data using various
emission models. cemekl: a thermal
plasma with power law temperature distribution -- a metal abundance of
solar (Z=1) and variable abundance are shown; cevmkl -- as cemekl but
the abundance of each element is allowed to vary -- a metal abundance
consistent with CNO processed composition, and when allowed to vary.
An absorption component was included in each model.}
\label{fits}
\end{table}

We show the X-ray spectrum of each of our 4 sources in Figure
\ref{spec}.  We then fitted various absorbed thermal plasma models to
the data using {\tt XSPEC}. With the exception of CR Boo, the fits to
an absorbed single temperature thermal plasma model were poor. We then
fitted each spectrum using an absorbed multi-temperature thermal
plasma model in which the emission measures follow a power-law
distribution ({\tt cemekl} in {\tt XSPEC}). Apart from CR Boo, which
gave a reasonable fit, the fits were also poor; this is similar to the
result when we allowed the overall metal abundance to vary (Table
\ref{fits}). We note that the spectra of CR Boo and HP Lib show
evidence of Fe K$\alpha$ emission, indicative of the presence of iron,
where as there was no such evidence in the spectra of AM CVn and HP
Lib. The strong emission features in GP Com are probably due to NVII
(near 0.5keV) and NeX (near 1keV). Therefore our X-ray spectral
fitting indicates that all our targets have non-solar He, C, N, O
abundances. The {\xmm} RGS observations of GP Com also show abundances
which are very different to solar values (Strohmayer 2004).

The abundance determinations can be understood as the donor star being
the remnant of a hydrogen exhausted stellar core. The high N abundance
suggests that the burning occurred at temperatures hot enough for the
CN(O) cycle to be the dominant burning process (ie $T$ \gtae $1.5
\times 10^7$, Bethe 1939).  In that case almost all the C (and at
higher temperatures O) is transformed into N (eg Caughlan \& Fowler
1962).

Using the stellar evolution code of Pols et al (1995) we determined
abundances of He=3.5, C=0.04, N=12.5, O=0.09 (all relative to solar)
for relatively low temperature CNO processed material, as appropriate
for relatively low-mass stars. Other elements were fixed at solar (we
used {\tt cevmkl} in {\tt XSPEC}). Apart from GP Com, all sources gave
fits which were significantly improved (Table \ref{fits}). We then
allowed the element abundances of He, C, N \& O to vary and found that
the fits to the spectra of GP Com and AM CVn were significantly
improved, while HP Lib and CR Boo were not. Both AM CVn and GP Com
show a significant enhancement of N compared to that expected from CNO
processed material (Table \ref{bestfits}).

\begin{figure*}
\begin{center}
\setlength{\unitlength}{1cm}
\begin{picture}(16,11)
\put(-0.5,-0.5){\includegraphics{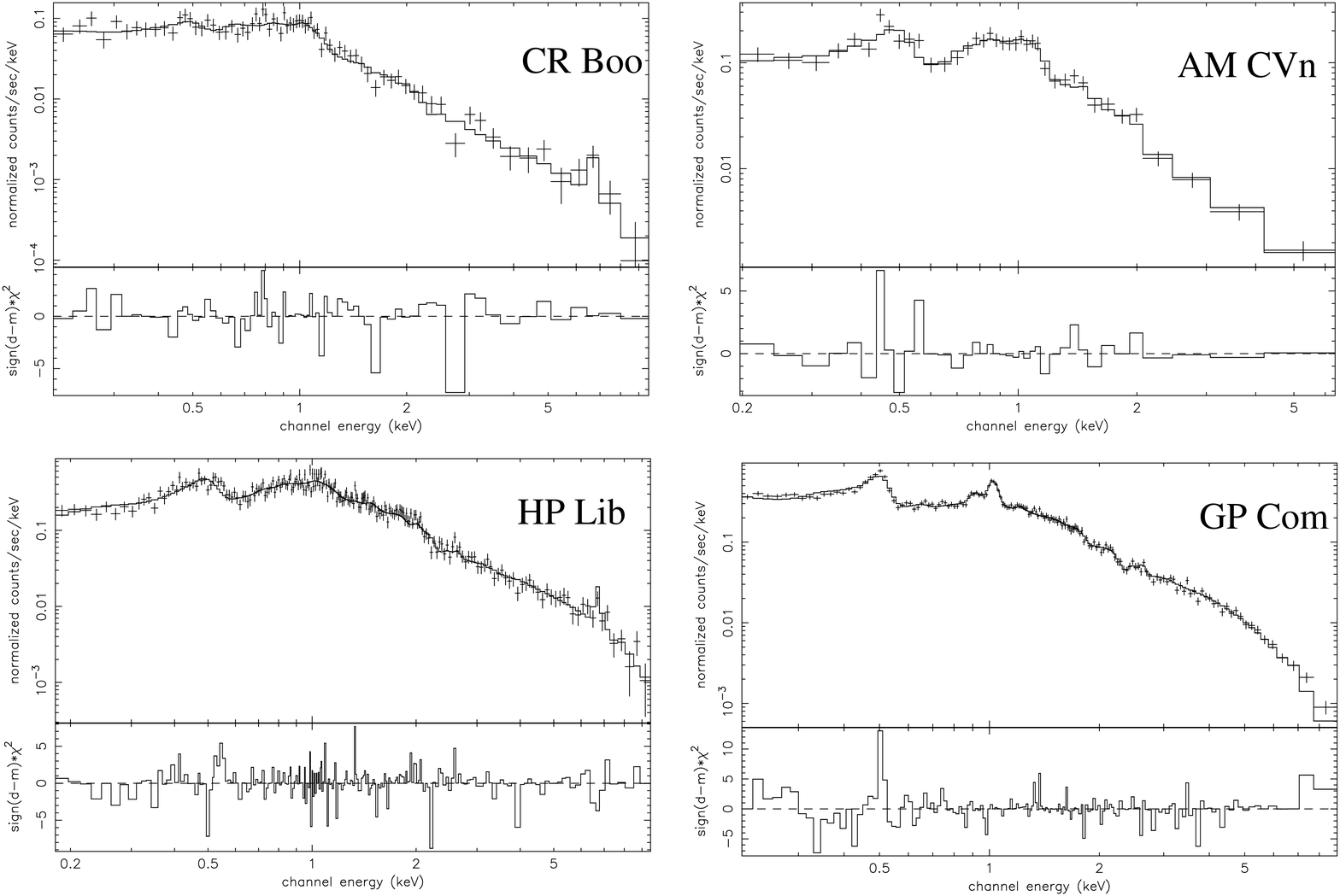}}
\end{picture}
\end{center}
\caption{The EPIC spectra of CR Boo, AM CVn, HP Lib \& GP Com together
with the best fit models overlayed.}
\label{spec}
\end{figure*}

In the case of GP Com, although the initial fit indicated an
enhancement of C relative to CNO, a more complete analysis revealed
that it was only the MOS1 spectrum that required C. Strohmayer (2004)
did not find evidence for the presence of C in his analysis of the
same data. To investigate this further, we did the same fitting
procedure as Strohmayer: this involved setting the abundances of
elements heavier than He to zero then allowing them to vary one by
one. If they made no significant improvement they were then fixed at
zero. This gave similar results to the previous result with the
exception of C which was not found to be required in the fit
(consistent with Strohmayer 2004). Using this approach we find best
fit abundance values of O=0.6$_{-0.2}^{+1.4}$, Ne=3.4$_{-1.5}^{+9.6}$,
S=0.9$_{-0.3}^{+3.2}$, Fe=0.1$_{-0.1}^{+0.6}$,
Ni=1.5$_{-1.0}^{+6.8}$. We conclude that there is not a significant
amount of C in the X-ray spectrum of GP Com.

We show the best fit spectral parameters in Table \ref{bestfits} along
with their confidence intervals. We also show the observed X-ray flux
in the 0.15--10keV energy band and the unabsorbed, bolometric X-ray
flux inferred for each source. We determine the bolometric X-ray
luminosities from recent parallax distance estimates - AM CVn 235 pc
(Dahn, referred to in Nelemans, Yungelson \& Portegies Zwart 2004); GP
Com 70 pc (Thortensen 2003). We find that both these sources have
X-ray luminosities $\sim4\times10^{30}$ erg/s.

How do these X-ray luminosities compare with the UV luminosity?  We
have one UV photometric point for AM CVn, and two for GP Com. This
makes it difficult to determine UV fluxes. However, our UV flux
measurements (shown in Table \ref{uvflux} and which were determined by
taking the count rate to flux conversion appropriate to white dwarfs
from the {\xmm} ESA web site) are within a factor of 2 compared to the
{\sl IUE} flux measurements (obtained from the {\sl IUE} archive at
STScI). We therefore integrated the {\sl IUE} spectra of AM CVn and GP
Com from the archive between 1100--3400\AA. Assuming the same
distances as before we find UV luminosities of 1.6$\times10^{33}$
erg/s and 1.3$\times10^{31}$ erg/s for AM CVn and GP Com
respectively. While the UV luminosity of GP Com is a factor of 3
greater than its X-ray luminosity, in AM CVn the UV luminosity is
nearly 3 orders of magnitude greater than its X-ray luminosity. We
discuss this further in \S \ref{discuss}.

\begin{table}
\begin{center}
\begin{tabular}{lrrrrrrrr}
\hline
Source & UV Flux (\ergscm \AA$^{-1}$)\\
\hline
AM CVn & 4.18$\times10^{-14}$ \\
CR Boo & 1.54$\times10^{-14}$ \\
HP Lib & 5.72$\times10^{-14}$ \\
GP Com & 5.32$\times10^{-15}$ \\
\hline
\end{tabular}
\end{center}
\caption{The inferred flux in the OM UVW1 filter assuming a conversion
from count rate to flux applicable to white dwarfs.}
\label{uvflux}
\end{table}

\section{Comparing the properties of AM CVn systems with other CVs}

The X-ray luminosities of AM CVn and GP Com are lower than most types
of hydrogen accreting CVs, which show $L_{X}\sim10^{31-33}$
\ergss. The diskless AM CVn systems, RX J1914+24 and RX J0806+15, show
very different X-ray spectra, showing prominent soft blackbody spectra
(Ramsay et al 2005).

\begin{table*}
\begin{center}
\begin{tabular}{lrrrrrrrr}
\hline
Source & $N_{H} $ & $\alpha$ & $T_{max}$ & Z & $F_{x,o}$ &
$F_{x,u}$ & $L_{X}$ \\
       & $\times10^{20}$ & & (keV) & (solar) & \ergss & \ergss & erg/s \\
       & \pcmsq & &  &  & cm$^{-2}$ & cm$^{-2}$ &  \\
\hline
AM CVn & 5.4$^{+3.5}_{-2.5}$ & 1.05$^{+0.42}_{-0.15}$ & 4.6$\pm$1.0 & 
30$^{+14}_{-10}$ (N) & 3.72$\times10^{-13}$ & 6.39$\times10^{-13}$ & 
4.2$\times10^{30}$ \\
CR Boo & 2.4$^{+1.4}_{-1.3}$ & 1.0$^{+0.2}_{-0.4}$ & 8.2$^{+4.0}_{-2.3}$ & 
14$^{+9}_{-7}$ (N) & 2.39$\times10^{-13}$ & 3.79$\times10^{-13}$ & &  \\
HP Lib & 13.3$^{+4.0}_{-2.4}$ & 0.8$^{+0.3}_{-0.3}$ &
8.4$^{+2.4}_{-1.6}$  &  
19$^{+10}_{-8}$ (N) & 1.59$\times10^{-12}$ & 3.10$\times10^{-12}$ & \\
%  & & & & 0.06$^{+0.24}_{-0.06}$ (O)  & & & \\ 
GP Com & 0.0$^{+0.1}$ & 0.86$^{+0.16}_{-0.02}$ &
8.0$\pm0.7$  & 30$^{+60}_{-15}$ (N)  & 5.50$\times10^{-12}$ & 
7.76$\times10^{-12}$ & 4.5$\times10^{30}$ \\ 
%  & & & & 0.6$^{+1.4}_{-0.2}$ (O)  & & &\\ 
%  & & & & 3.4$^{+9.6}_{-1.5}$ (Ne)  & & &\\ 
%  & & & & 0.9$^{+3.2}_{-0.3}$ (S)  & & &\\ 
%  & & & & 0.11$^{+0.61}_{-0.07}$ (Fe)  & & &\\ 
%  & & & & 1.5$^{+6.8}_{-1.0}$ (Ni)  & & &\\   
\hline
\end{tabular}
\end{center}
\caption{The spectral parameters derived from fitting an absorbed
multi-temperature thermal plasma model with variable metal abundance,
Z, to the {\xmm} EPIC data. The slope of the power law distribution of
temperature, $\alpha$, and the maximum temperature, $T_{max}$ of the
plasma are shown.  We show the observed X-ray flux in the 0.15-10keV
band, $F_{x,o}$, the unabsorbed, bolometric X-ray flux, $F_{x,u}$, the
bolometric X-ray luminosity, $L_{X}$.}
\label{bestfits}
\end{table*}

\begin{figure}
\begin{center}
\setlength{\unitlength}{1cm}
\begin{picture}(6,5.5)
\put(-1.,0){\includegraphics{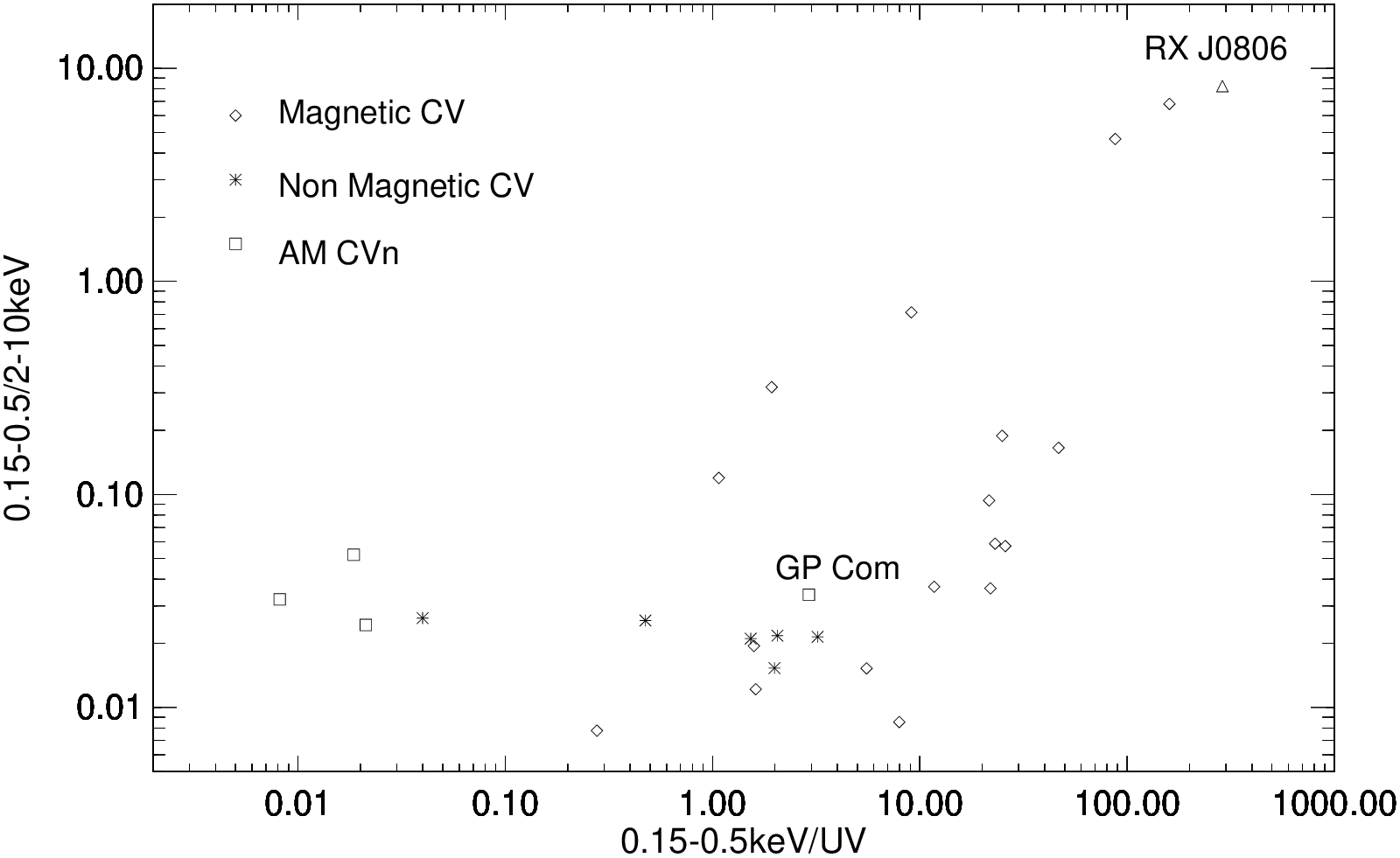}}
\end{picture}
\end{center}
\caption{The colours for the AM CVn systems in this paper;
non-magnetic CVs and the strongly magnetic CVs in the
0.15-0.5/2-10keV, 0.15-0.5/UV plane.}
\label{xraycols}
\end{figure}

To extend the comparison further we compare the relative flux in
soft/hard X-rays with the relative flux in soft X-rays/UV for various
varieties of CVs.  We determined the flux in the 0.15--0.5keV and the
2--10keV band and normalised them to give the flux in \ergscm
\AA$^{-1}$. We corrected the fluxes for the extinction that was
determined from the X-ray spectrum for each source individually. We
plot the ratio of our AM CVn sample in Figure \ref{xraycols} in the
0.15--0.5/UVW1, 0.15-0.5/2--10keV plane. We also show the strongly
magnetic CVs (the polars) taken from the sample of Ramsay \& Cropper
(2004) and the non-magnetic CVs taken from the {\xmm} public
archive. We find that the soft/hard X-ray ratio of the AM CVn systems
are similar to that found for the non-magnetic CVs, while most of the
polars show a higher soft/hard ratio since they have a significant
soft X-ray component due to re-processing of the hard component. On
the other hand, for the AM CVn systems the soft X-ray/UV ratio is very
low (apart from GP Com) which indicates that compared to either the
non-magnetic and magnetic sample, the AM CVn systems have a strong UV
component, or a weak soft X-ray component. In contrast, RX J0806+15 is
at an extreme end of the diagram, which is expected for a strong soft
X-ray source with both weak UV and hard X-ray emission. (The
interstellar extinction to RX J1914+24 is so high that the UV emission
is not detectable).

\section{Discussion}
\label{discuss}

\subsection{Light curves}

The X-ray and UV light curves of the 4 systems in our sample are very
different. While the sources show variability in their X-ray
intensity, there is only marginal evidence for coherent
modulation. This is similar to that of non-magnetic CVs, most of which
show no evidence for significant X-ray modulations at periods shorter
than a few hours (eg Baskill, Wheatley \& Osborne 2005). Those few
non-magnetic CVs which {\sl have} shown evidence, eg OY Car (Ramsay et
al 2001), have been associated with the spin period of the white
dwarf. This implies that the accretion flow is controlled by the
magnetic field of the white dwarf as it nears the white dwarf
suggesting a significant magnetic field strength (of order 10$^{5}$
G). Our observations suggest that the accreting white dwarfs in the
disc accreting AM CVn systems presented here do not have significant
magnetic fields. 

In contrast, the UV observations of our sample shows remarkably
heterogeneous behaviour. CR Boo shows a significant increase in
intensity over the 5.5 hr observation -- the cause of this variation
is not clear. In HP Lib, there is a strong coherent modulation. It is
similar to that of the UV modulation in the intermediate polar FO Aqr
(Evans et al 2004). In these systems the accreting white dwarfs
disrupts the formation of a disc close to the white dwarf and accretes
via a disc-fed accretion `curtain'. However, unlike FO Aqr, HP Lib
shows no strong evidence for a coherent modulation in X-rays. In AM
CVn we detect a period at 996 sec: this period is not significantly
different previously reported periods at 1011.4 and 1014 sec.

\subsection{Accretion Rates}

We determine the accretion luminosity by simply summing up the X-ray
and UV luminosity. In practice some fraction of the UV luminosity will
originate from the disc and/or the white dwarf. Further, it is not
clear how the lack of hydrogen will affect the optical thickness of
the X-ray and UV emitting regions. We therefore find
$L_{tot}=1.6\times10^{33}$ \ergss and 1.7$\times10^{31}$ \ergss for AM
CVn and GP Com respectively. We determine the accretion rate from
$L_{tot}=GM\dot{M}/R$ (ie the sum of the accretion disc emission and
the emission from the region close to the accreting object). We
therefore obtain $\dot{M}=1.0\times10^{16}$ g/s (=1.6$\times10^{-10}$
\Msun/yr) and 1.1$\times10^{14}$ g/s (=1.7$\times10^{-12}$ \Msun/yr)
for AM CVn and GP Com respectively for a 0.6\Msun accreting white
dwarf. These are lower limits to the accretion rate since we did not
include UV emission outside the range 1100--3400\AA (\S 4) and we did
not include optical emission.

How do the above values compare with other estimates of the accretion
rate in these objects? Nasser et al (2001) and Nagel et al (2004)
modelled optical spectra of several AM CVn systems using NLTE models
and both determined an accretion rate of 3$\times10^{-9}$ \Msun/yr for
AM CVn.  This is consistent with theoretical work which
has predicted an accretion rate of 1$\times10^{-8}-4\times10^{-10}$
\Msun/yr (Deloye, Bilstein \& Nelemans 2005). This suggests that for
AM CVn, we do not observe a significant proportion of the accretion
energy. One possible solution is that the material is lost in the form
a wind (Solheim, Provencal \& Sion 1997).

\subsection{The sources of emission in AM CVn systems}

The AM CVn systems in our survey all show X-ray spectra which are best
modelled using a multi-temperature emission model and a strong UV
component. The systems are not only hydrogen deficient but also show
non-solar metalicities. Figure \ref{xraycols} indicates that compared
to hydrogen donor CVs, the disk-accreting AM CVn systems (with the
exception of the longer period system GP Com) show low soft X-ray/UV
colours, but similar soft X-ray/hard X-ray ratios. This implies that
the UV component plays a more dominant role in disk accreting AM CVns
or that the X-ray component is significantly less prominent than
expected.

In non-magnetic CVs, the soft X-ray emission is expected to be
dominated by boundary layer emission as opposed to emission from the
extended disc. At high mass transfer rates, this boundary layer is
optically thick and emits chiefly in the UV regime (eg Popham \&
Narayan 1995), while at low mass transfer rates the optically thin
layer can contribute to harder X-ray emission. The lack of hydrogen in
AM CVn systems will not only affect the properties of their helium
dominated accretion disks, but also their boundary layer. Although it
is not obvious as to how this will affect the overall spectrum, it is
expected that given the higher mass transfer rates, a strong UV
component suggests this optically thick boundary layer picture for the
UV/soft X-ray emission also applies to the short period AM CVn
systems. This is consistent with observations of hydrogen accreting
dwarf novae CVs undergoing an outburst (ie high accretion rates),
which show an increase in their UV flux and a decrease in soft X-rays
(eg Wheatley et al 1996). We also note that the stability of the disc
is influenced by the metal abundance of the disc (eg Menou, Perna \&
Hernquist 2002).

\subsection{The element abundances}

Our X-ray spectral fits showed that a good fit was obtained for HP Lib
and CR Boo when we assumed a metal abundance consistent with that
expected from CNO processed material. In the case of AM CVn a slightly
higher abundance of N gave a significantly better fit (99.7 percent
significance). In the case of GP Com, the abundance of N and O is
enhanced compared to low temperature CNO reprocessed material, with
evidence of significant amounts of Ne, S, Fe and Ni (\S 4).

How do these results compare with optical results? Marsh, Horne \&
Rosen (1991) found one oxygen atom for every 50 nitrogen atoms.
Taking solar abundance values from Anders \& Grevesse (1989) we find a
higher number, one in $\sim$6. We determine a ratio of log(N/Fe)=2.8
compared to $>$3.5 from Marsh, Horne \& Rosen (1991). The X-ray fits
were not very sensitive to the abundance of helium.

Some difference can be expected between the optical and X-ray results
because of our ignorance of the exact physical conditions in these
systems. The optical abundances were based upon single temperature LTE
models for atoms of very different excitation and ionisation energy,
namely neutral helium and nitrogen. The overall abundance of nitrogen
from the optical data is therefore uncertain. It is less obvious that
the optical N/Fe ratio should be far wrong given the comparable
ionisation energies of NI and FeII. However, it is possible that high
optical depths in the NI lines in the LTE model lead to an
overestimate of this ratio, assuming that there was in fact some
unaccounted-for broadening mechanism which could raise the line
strength for a given abundance.

\section{Conclusions}

We have presented observations of 4 disk accreting AM CVn
systems. Their X-ray and UV intensity variations are heterogeneous,
with only HP Lib showing a very clear coherent modulation in its UV
light curve. Their X-ray spectra are best fitted with highly non-solar
abundances, with CR Boo and HP Lib showing abundances consistent with
that expected from low temperature CNO reprocessed material. AM CVn
and GP Com show an enhancement of nitrogen, while GP Com shows an
enhancement of additional elements.  By comparing the abundances
determined using X-ray, UV and optical methods in detail, it will be
possible to gain a better understanding of the underlying physical
nature of their emission sites. Moreover, once we have a better
understanding of their abundances we can in principal determine the
systems evolutionary history. A large fraction of the accretion
luminosity is emitted in the UV.

\begin{acknowledgements}

This paper is based on observations obtained using {\sl XMM-Newton},
an ESA science mission with instruments and contributions directly
funded by ESA Member States and the USA (NASA).  PH is supported by
the Academy of Finland and DS acknowledges a Smithsonian Astrophysical
Observatory Clay Fellowship. We thank the referee, Jan-Erik Solheim,
for useful comments which helped improve the paper.

\end{acknowledgements}


\begin{thebibliography}{99}

%\bibitem{}Allen, C. W., 1976, Third Edition, Athlone Press
\bibitem{}Anders, E., Grevesse, N., 1989, Geochimica et Cosmochimica
Acta, 53, 197
\bibitem{}Baskill, D. S., Wheatley, P. J., Osborne, J. P., 2005, MNRAS, 
357, 626
\bibitem{}Bethe, H. A., 1939, Physical Review, 55, 434
\bibitem{}Caughlan, G. R. \& Fowler, W. A., 1962, ApJ, 136, 453
\bibitem{}Deloye, C. J., Bilstein, L., \& Nelemans, G., 2005, accepted ApJ,
astroph/0501577 
%\bibitem{}Cropper, M., Ramsay, G., Wu, K., Hakala, P., 2004, In Proc Third
%Workshop on magnetic CVs, Cape Town, astro-ph/0302240
%\bibitem{}Espaillat, C., Patterson, J., Warner, B., Woudt, P.,
%accepted, PASP, astro-ph/0412068
\bibitem{}Evans, P. A., Hellier, C., Ramsay, G., Cropper, M., 2004,
MNRAS, 349, 715
%\bibitem{}Frank, J., King, A., Raine, D., 2002, Accretion Power in
%Astrophysics, Third Edition, Cambridge University Press
%\bibitem{}Hakala, P., Ramsay, G., 2004, A\&A, 416, 1047
\bibitem{}Hakala, P., Ramsay, G., Wheatley, P., Harlaftis, E. T.,
Papadimitriou, C., 2004, A\&A, 420, 273
%\bibitem{}Iben, I., Tutukov, A. V., 1991, ApJ, 370, 615
%\bibitem{}Israel, G. L., et al, 2003, ApJ, 598, 492
\bibitem{}Kato, T., Nogami, D., Baba, H., Hanson, G., Poyner, G.,
2000, MNRAS, 315, 140
\bibitem{}Marsh, T., Horne, K., Rosen, S., 1991, MNRAS, 366, 535
\bibitem{}Mason K. O., et al 2001, A\&A, 365, L36
\bibitem{}Menou, K., Perna, R.,  \& Hernquist, L., 2002, ApJ, 564, L81
%\bibitem{}Nather, R. E., Robinson, E. L., Stover, R. J., 1981, ApJ,
%244, 269
\bibitem{}Nagel, T., Dreizler, S., Rauch, T., Werner, K., 2004, A\&A,
428, 109
\bibitem{}Nasser, M. R., Solheim, J.-E., Semionoff, D. A., 2001, A\&A,
373, 333
\bibitem{}Nelemans, G., In Proc `{\sl The astrophysics of cataclysmic 
variables and related objects}', ASP Conf. Ser., Eds. Hameury, J. M. \& 
Lasota, J. P., 2004, astro-ph/0409676 
%\bibitem{}Nelemans, G., Steeghs, D., Groot, P., 2001, MNRAS, 326, 621
%\bibitem{}Nelemans, G., Portegies Zwart, S. F., Verbunt, F.,
%Yungelson, L R., 2001, A\&A, 368, 939
\bibitem{}Nelemans G., Yungelson L. R., Portegies Zwart S. F., 2004,
MNRAS, 349, 181
%\bibitem{}Pandel, D., Cordova, F. A., Howell, S. B., 2003, MNRAS, 336, 1049
%\bibitem{}Patterson, J., Halpern, J., Shambrook, A., 1993, ApJ, 419,
%803
\bibitem{}Patterson, J., et al, 2002, PASP, 114, 65
\bibitem{}Popham, R., Narayan, R., 1995, ApJ, 442, 337
\bibitem{}Pols, O. R., Tout, C. A., Eggleton, P. A, Han, Z., 1995,
MNRAS, 274, 964
\bibitem{}Provencal, J. L. et al, 1995, ApJ, 445, 927
\bibitem{}Provencal, J. L. et al, 1997, ApJ, 480, 383
\bibitem{}Ramsay, G., et al 2001, A\&A, 365, L288
\bibitem{}Ramsay, G., Cropper, M., 2004, MNRAS, 347, 497
%\bibitem{}Ramsay, G., Cropper, M., Wu, K., Mason, K. O., Cordova,
%F. A., Priedhorsky, W., 2004, MNRAS, 350, 1373
\bibitem{}Ramsay, G., Hakala, P., Wu, K., Cropper, M., Mason, K.,
Cordova, F. A., Priedhorsky, W., 2005, MNRAS, 357, 49
%\bibitem{}Solheim, J. E., et al, 1998, A\&A, 332, 939
\bibitem{}Solheim, J. E., Provencal, J. L., Sion, E. M., 1997, In {\sl
White dwarfs}, Proceedings of the 10th European Workshop on White
Dwarfs, Ed. J. Isern, M. Hernanz, and E. Gracia-Berro;
Kluwer Academic Publishers, Dordrecht, Astro \& Space 
Sci Lib, Vol. 214, p.337
%\bibitem{}Steeghs, D., et al, in prep
\bibitem{}Strohmayer, T. E., 2004, ApJ, 608, L53
\bibitem{}Str\"{u}der L., et al, 2001, 365, L18
\bibitem{}Thorstensen J R., 2003, AJ, 126, 3017
\bibitem{}Turner M., et al 2001, A\&A, 365, L27
%\bibitem{}van Teeseling, A., Verbunt, F., 1994, A\&A, 292, 519
\bibitem{}van Teeseling, A., Beuermann, K., Verbunt, F., 1996, A\&A,
315, 467
%\bibitem{}Verbunt, F., Bunk, W. H., Ritter, H., Pfeffermann, E., 1997,
%A\&A, 327, 602
\bibitem{}Ulla, A., 1995, A\&A, 301, 469
\bibitem{}Warner B., 1995, Cataclysmic variable stars, Cambridge
Univ. Press, Cambridge
%\bibitem{}Warner, B., Woudt, P. A., 2002, PASP, 114, 129
\bibitem{}Wheatley, P. J., Verbunt, F., Belloni, T., Watson, M. G., 
Naylor, T., Ishida, M., Duck, S. R., Pfeffermann, E., 1996, A\&A, 307, 137
%\bibitem{}Wood, M. A., Winget, D. E., Nather, R. E., Hessman, F. V.,
%Liebert, J., Kurtz, D. W., Wesemael, F., Wegner, G., 1987, ApJ, 313, 757

\end{thebibliography}
\end{document}